# RodFIter: Attitude Reconstruction from Inertial Measurement by Functional Iteration

Yuanxin Wu, *Senior Member, IEEE*

*Abstract*—Rigid motion computation or estimation is a cornerstone in numerous fields. Attitude computation can be achieved by integrating the angular velocity measured by gyroscopes, the accuracy of which is crucially important for the dead-reckoning inertial navigation. The state-of-the-art attitude algorithms have unexceptionally relied on the simplified differential equation of the rotation vector to obtain the attitude. This paper proposes a Functional Iteration technique with the Rodrigues vector (named the RodFIter method) to analytically reconstruct the attitude from gyroscope measurements. The RodFIter method is provably exact in reconstructing the incremental attitude as long as the angular velocity is exact. Notably, the Rodrigues vector is analytically obtained and can be used to update the attitude over the considered time interval. The proposed method gives birth to an ultimate attitude algorithm scheme that can be naturally extended to the general rigid motion computation. It is extensively evaluated under the attitude coning motion and compares favorably in accuracy with the mainstream attitude algorithms. This work is believed having eliminated the long-standing theoretical barrier in exact motion integration from inertial measurements.

## I. INTRODUCTION

Rigid motion is the basic transformation of objects in the Euclidean space. Three-dimensional rigid motion computation or estimation is a cornerstone in vast fields, such as physics, robotics, guidance and navigation, mechanics and computer vision, to name but only a few. In contrast to the translation motion, such as position and linear velocity, the orientation or attitude cannot be directly measured in any means. Instead, the attitude is either computed by integrating the angular velocity or estimated by matching some vectors measured in respective frames [4]. The former method is preferred in many areas like the GPS-denied navigation, as it is self-contained and does not need any external information for considerable time duration [6, 7]. However, the angular velocity measurement, for example by gyroscopes, is inevitably contaminated by errors, which will lead to unbounded error accumulation through time integration. It has been believed so far that even if the angular velocity measurement is perfect and error-free, there still exists an insurmountable fact that prevents us from practically obtaining the exact attitude, namely, the non-commutativity attribute of finite rotations [9]. That is to say, switching the order of consecutive rotations normally produces different attitude. So far, approximation has to be introduced into the attitude computation.

Attitude computation is crucially important for the dead-reckoning inertial navigation, so great endeavors have been devoted to improving the attitude computation accuracy as much as possible, see e.g. [3, 10]. Table I lists the landmark papers so far on approximate attitude computation using the rotation vector. The structure of the state-of-the-art or mainstream algorithms was established in early 1970s, pioneered by Jordan [1] and Bortz [9], which exclusively relies on the rotation vector to represent the incremental attitude and then update the current attitude in terms of quaternion or direction cosine matrix [1, 9, 11]. As the rotation vector rate is nontrivially related to the gyroscope-measured angular velocity, it has to be much simplified so as to be handled by practical approximation methods [3, 11]. The attitude algorithms need several consecutive measurements (called samples) to approximate the incremental rotation vector and thus cannot update the current attitude until the last measurement comes in. It is not a big problem when attitude is what we are only concerned about, but it is a different story if the subsequent computation step needs the current attitude as an input. The latter case is what happens to the inertial navigation system [12-14]. To better approximate the attitude, more angular velocity samples are preferred, but the velocity or position computation is awaiting the attitude result to come out as soon as possible, also for the sake of better approximation. In addition, the state-of-the-art attitude algorithms are all designed using some special form of motion, e.g., the coning motion, as a performance optimization criterion [2, 5, 15] and may likely be suboptimal for other kinds of motions.

The paper proposes a functional iteration technique to analytically reconstruct the attitude from gyroscope measurements by way of the Rodrigues vector (named as the RodFIter method hereafter). The Rodrigues vector is used instead of the traditional rotation vector to represent the incremental attitude as it has a much simpler rate equation that enables exact integration iteration in terms of finite-order polynomials. The contribution of this paper is multiple-folded. Firstly, the RodFIter method gives birth to a brand-new and ultimate attitude algorithm scheme that can be naturally extended to the general rigid motion computation. Secondly,

This work was supported in part by National Natural Science Foundation of China (61422311, 61673263), Joint Fund of China Ministry of Education (6141A02022309) and Hunan Provincial Natural Science Foundation of China (2015JJ1021). A short version was presented at the symposium on Inertial Sensors and Systems (ISS), Karlsruhe, Germany, 2017.

Authors' address: Shanghai Key Laboratory of Navigation and Location-based Services, School of Electronic Information and Electrical Engineering, Shanghai Jiao Tong University, Shanghai, China, 200240, E-mail: (yuanx_wu@hotmail.com).

Table I. Landmark works on Approximate Attitude Computation Using Rotation Vector

| Landmark Works | Approximation of Rotation Vector Rate[#] |
|---|---|
| [1] by Jordan, 1969<br>[2] by Ignagni, 1990<br>[3] by Savage, 1998 | $\dot{\mathbf{g}} = \boldsymbol{\omega} + \frac{1}{2}\boldsymbol{\theta}\times\boldsymbol{\omega}, \quad \dot{\boldsymbol{\theta}} = \boldsymbol{\omega}$ |
| [5] by Miller, 1983 | $\dot{\mathbf{g}} = \boldsymbol{\omega} + \frac{1}{2}\mathbf{g}\times\boldsymbol{\omega}$ |
| [8] by Wang, 2015 | $\dot{\mathbf{g}} = \boldsymbol{\omega} + \frac{1}{2}\left(\boldsymbol{\theta} + \frac{1}{2}\int\boldsymbol{\theta}\times\boldsymbol{\omega}\right)\times\boldsymbol{\omega} + \frac{1}{12}\boldsymbol{\theta}\times(\boldsymbol{\theta}\times\boldsymbol{\omega})$ |

[#]: Please refer to Section II for symbol definitions

the new attitude algorithm is theoretically exact as long as the angular velocity is perfect. The incremental Rodrigues vector is obtained in the analytic form and thus the RodFIter method is able to produce attitude over the whole considered time interval. Thirdly, the RodFIter method does not depend on any special form of motion, unlike the mainstream attitude algorithms being designed with the coning motion as the criterion, and can be optimally applied to any kinds of motions.

The paper is organized as follows. Section II revisits the rotation vector and the Rodrigues vector, and their differential equations. Section III presents the angular velocity fitting of angular velocity/increment measurements by the Chebyshev polynomial and implements the Rodrigues vector construction by way of an iteration process of functional integration. The convergence of the iteration process is provably guaranteed for cases of both perfect and erroneous measurements. Section IV discusses the natural extension of the proposed RodFIter method to the general rigid motion. Section V evaluates the proposed attitude algorithm under the coning motion and compares with the mainstream algorithms. Section VI concludes the paper.

## II. GENERALIZED KINEMATIC VECTOR

### A. Definition and Its Relation to Attitude

Among a number of parameters to represent the attitude, quaternion is the most preferred as it has the minimum number of elements and is free of singularity. Let $\mathbf{q}$ encode the attitude quaternion of the body frame relative to some reference frame. The Euler theorem states that any composition of finite rotations can be realized by one equivalent rotation about some fixed axis [16, 17]. Denote the unit vector of the fixed axis and the magnitude of the rotation by $\mathbf{e}$ and $\alpha$ respectively, the quaternion can be expressed as

$$\mathbf{q} = \cos\frac{\alpha}{2} + \mathbf{e}\sin\frac{\alpha}{2} \qquad (1)$$

The quaternion differential equation is given by

$$2\dot{\mathbf{q}} = \mathbf{q} \circ \boldsymbol{\omega} \qquad (2)$$

where $\boldsymbol{\omega}$ is the angular velocity vector expressed in the body frame and $\circ$ denotes the quaternion multiplication [6].

Alternatively, the generalized kinematic vector can be used to describe the equivalent rotation [4, 17]

$$\mathbf{g} = f(\alpha)\mathbf{e} \qquad (3)$$

where $f(\cdot)$ denotes a scalar function. The kinematic vector becomes the rotation vector when $f(\alpha) = \alpha$, the Rodrigues vector when $f(\alpha) = 2\tan\left(\frac{\alpha}{2}\right)$. We next derive the differential equation of the generalized kinematic vector using the property of quaternion [4, 17].

Substitute (1) into (2) and decompose the resultant into the scalar and vector parts

$$\begin{aligned}\dot{\alpha} &= \boldsymbol{\omega}\cdot\mathbf{e} \\ \dot{\mathbf{e}} &= \frac{1}{2}\mathbf{e}\times\boldsymbol{\omega} + \frac{1}{2}\mathbf{e}\times(\mathbf{e}\times\boldsymbol{\omega})\cot\frac{\alpha}{2}\end{aligned} \qquad (4)$$

where $\times$ denotes the vector cross product or the skew-symmetric matrix formed according to the definition of the vector cross product. Differentiate the generalized kinematic vector (3),

$$\dot{\mathbf{g}} = \frac{\partial f}{\partial \alpha}\dot{\alpha}\mathbf{e} + f(\alpha)\dot{\mathbf{e}} \triangleq f'_\alpha(\alpha)\dot{\alpha}\mathbf{e} + f(\alpha)\dot{\mathbf{e}} \qquad (5)$$

Substituting (4) into (5), the differential equation of the generalized kinematic vector is

$$\dot{\mathbf{g}} = f'_\alpha(\alpha)\boldsymbol{\omega} + \frac{1}{2}\mathbf{g}\times\boldsymbol{\omega} + \frac{1}{f^2(\alpha)}\left(f'_\alpha(\alpha) - \frac{1}{2}f(\alpha)\cot\frac{\alpha}{2}\right)\mathbf{g}\times(\mathbf{g}\times\boldsymbol{\omega}) \qquad (6)$$

The last two terms are collectively referred to as the noncommutativity rate vector [9], which clearly shows that the current kinematic vector is an *in-order* accumulating result of the history kinematic vector and the angular velocity up to current time [18].

From (2), the angular velocity can be obtained as

$$\boldsymbol{\omega} = 2\mathbf{q}^* \circ \dot{\mathbf{q}} = \dot{\alpha}\mathbf{e} + \sin\alpha\,\dot{\mathbf{e}} - (1-\cos\alpha)\mathbf{e}\times\dot{\mathbf{e}} \qquad (7)$$

where $\mathbf{q}^*$ is the conjugate quaternion. Using (3) and (4), it is rewritten as

$$\boldsymbol{\omega} = \frac{1}{f'_\alpha(\alpha)}\dot{\mathbf{g}} - \frac{1-\cos\alpha}{f^2(\alpha)}\mathbf{g}\times\dot{\mathbf{g}} + \frac{1}{f^2(\alpha)}\left(\frac{1}{f'_\alpha(\alpha)} - \frac{\sin\alpha}{f(\alpha)}\right)\mathbf{g}\times(\mathbf{g}\times\dot{\mathbf{g}}) \qquad (8)$$

Note that (8) is the reverse formula of (6) that expresses the angular velocity vector as a function of the generalized kinematic vector and its derivative.

### B. Rotation Vector and Rodrigues Vector

The rotation vector is defined as $\mathbf{g} = \alpha\mathbf{e}$ and the quaternion

can be obtained as

$$\mathbf{q} = \cos\frac{|\mathbf{g}|}{2} + \frac{\mathbf{g}}{|\mathbf{g}|}\sin\frac{|\mathbf{g}|}{2} \qquad (9)$$

where $|\cdot|$ means the magnitude of a vector. The differential equation of the rotation vector and the reverse formula are

$$\dot{\mathbf{g}} = \boldsymbol{\omega} + \frac{1}{2}\mathbf{g}\times\boldsymbol{\omega} + \frac{1}{|\mathbf{g}|^2}\left(1 - \frac{|\mathbf{g}|\sin|\mathbf{g}|}{2(1-\cos|\mathbf{g}|)}\right)\mathbf{g}\times(\mathbf{g}\times\boldsymbol{\omega}) \quad (10)$$

$$\boldsymbol{\omega} = \dot{\mathbf{g}} - \frac{1-\cos|\mathbf{g}|}{|\mathbf{g}|^2}\mathbf{g}\times\dot{\mathbf{g}} + \frac{1}{|\mathbf{g}|^2}\left(1-\frac{\sin|\mathbf{g}|}{|\mathbf{g}|}\right)\mathbf{g}\times(\mathbf{g}\times\dot{\mathbf{g}}) \quad (11)$$

The rotation vector has a closed-form relation to the attitude matrix as

$$\mathbf{C} = \mathbf{I}_3 - \frac{\sin(|\mathbf{g}|)}{|\mathbf{g}|}\mathbf{g}\times + \frac{1-\cos(|\mathbf{g}|)}{|\mathbf{g}|^2}(\mathbf{g}\times)^2 \qquad (12)$$

where $\mathbf{C}$ denotes the attitude matrix of the body frame relative to the reference frame.

The Rodrigues vector is defined as $\mathbf{g} = 2\tan\left(\frac{\alpha}{2}\right)\mathbf{e}$ and the quaternion is obtained as

$$\mathbf{q} = \frac{2+\mathbf{g}}{\sqrt{4+|\mathbf{g}|^2}} \qquad (13)$$

The differential equation of the Rodrigues vector and the reverse formula are

$$\dot{\mathbf{g}} = \boldsymbol{\omega} + \frac{1}{2}\mathbf{g}\times\boldsymbol{\omega} + \frac{1}{4}\mathbf{g}(\mathbf{g}\cdot\boldsymbol{\omega}) \qquad (14)$$

$$\boldsymbol{\omega} = \frac{1}{4+|\mathbf{g}|^2}(4\dot{\mathbf{g}} - 2\mathbf{g}\times\dot{\mathbf{g}}) \qquad (15)$$

The Rodrigues vector has a closed-form relation to the attitude matrix as

$$\mathbf{C} = \mathbf{I}_3 - \frac{4}{4+|\mathbf{g}|^2}\mathbf{g}\times + \frac{2}{4+|\mathbf{g}|^2}(\mathbf{g}\times)^2 \qquad (16)$$

Equation (10) or (14) tells that the attitude cannot be simply computed by integrating the angular velocity, due to the existence of the noncommutativity rate vector. The Rodrigues vector has an apparently much simpler differential equation than the rotation vector does. The trigonometric functions in (10) and (11) do not exist in (14) and (15), owing to the special definition of the Rodrigues vector.

### C. Approximation in Rotation Vector

For the fixed-axis rotation, the rotation vector rate equation (10) (with the last two terms vanishing) reduces to $\dot{\mathbf{g}} = \boldsymbol{\omega}$ while the Rodrigues vector rate equation (14) (with the middle term vanishing) only reduces to $\dot{\mathbf{g}} = \boldsymbol{\omega} + \frac{1}{4}\mathbf{g}(\mathbf{g}\cdot\boldsymbol{\omega})$, so the attitude algorithms have exclusively been developed from the rotation vector, for instance, in the inertial navigation field [1, 3, 6, 7, 10]. For the general rotation, the rotation/Rodrigues vector appears on the right side of the nonlinear differential equation, to which there is no analytical solution. All the modern-day attitude algorithms, pioneered by Jordan [1] and Bortz [9], approximate the right-sided rotation vector by the integrated angular velocity and then (10) is simplified as [3, 6, 7, 10, 11]

$$\dot{\mathbf{g}} \approx \boldsymbol{\omega} + \frac{1}{2}\left(\int_0^t \boldsymbol{\omega}\,dt\right)\times\boldsymbol{\omega} + \frac{1}{12}\left(\int_0^t \boldsymbol{\omega}\,dt\right)\times\left(\left(\int_0^t \boldsymbol{\omega}\,dt\right)\times\boldsymbol{\omega}\right)$$
(17)

where the trigonometric function $\frac{1}{|\mathbf{g}|^2}\left(1 - \frac{|\mathbf{g}|\sin|\mathbf{g}|}{2(1-\cos|\mathbf{g}|)}\right) \approx \frac{1}{12}$

if the rotation vector's magnitude $|\mathbf{g}|$ approaches zero. In fact, the attitude algorithms in practical inertial navigation systems have only considered the first two terms in (17) so as to alleviate the complexity of the algorithm design [11], and the last term has not been taken into account until quite recently [8]. An obsolete attempt was to solve the apparently linear rate equation of the attitude matrix in terms of three column vectors by the recursive Picard integration [19]. Besides the numerical integration approximation made therein, the adverse affect, incurred by the fact that the column vectors of the attitude matrix are of unity norm, was largely disregarded.

### III. ATTITUDE RECOVERY BY FUNCTIONAL ITERATION

In view of the simpler differential equation for the general rotation, the Rodrigues vector will be used instead to compute the attitude. The proposed algorithm builds on an iterative process and is proved to be convergent to the true attitude. Integrating (14), without the loss of generality, over the time interval $[0\ t]$,

$$\mathbf{g} = \int_0^t \left(\mathbf{I}_3 + \frac{1}{2}\mathbf{g}\times + \frac{1}{4}\mathbf{g}\mathbf{g}^T\right)\boldsymbol{\omega}\,dt \qquad (18)$$

#### A. Reconstructing Rodrigues Vector from Angular Velocity

Construct an iterative process to calculate the Rodrigues vector as such

$$\mathbf{g}_{j+1} = \int_0^t \left(\mathbf{I}_3 + \frac{1}{2}\mathbf{g}_j\times + \frac{1}{4}\mathbf{g}_j\mathbf{g}_j^T\right)\boldsymbol{\omega}\,dt \qquad (19)$$

with the initial function $\mathbf{g}_0 = 0$. For instance, the first two iterations yield

$$\mathbf{g}_1 = \int_0^t \boldsymbol{\omega}\,dt$$

$$\mathbf{g}_2 = \int_0^t \left(\mathbf{I}_3 + \frac{1}{2}\left(\int_0^t \boldsymbol{\omega}\,dt\right)\times + \frac{1}{4}\left(\int_0^t \boldsymbol{\omega}\,dt\right)\left(\int_0^t \boldsymbol{\omega}\,dt\right)^T\right)\boldsymbol{\omega}\,dt$$
(20)

*Theorem 1*: Given the true angular velocity function $\boldsymbol{\omega}$, the iterative process (19) converges to the true Rodrigues vector function over the interval when $\prod_{k=0}^{\infty}\left|\boldsymbol{\theta}_{t_k}\right|$ is bounded. (Definition of $\boldsymbol{\theta}_{t_k}$ is given below)

Proof. For the true angular velocity $\boldsymbol{\omega}$ over the time interval $[0\ t]$, there must exist a true Rodrigues vector $\mathbf{g}^*$ satisfying (18) and corresponding to the true attitude motion, that is, $\mathbf{g}^* = \int_0^t \left(\mathbf{I}_3 + \frac{1}{2}\mathbf{g}^*\times + \frac{1}{4}\mathbf{g}^*\mathbf{g}^{*T}\right)\boldsymbol{\omega}\,dt$. Define the error function $\boldsymbol{\varepsilon}_j \triangleq \mathbf{g}_j - \mathbf{g}^*$. Then the error function $\boldsymbol{\varepsilon}_j$ at any time

$t_j \in [0 \quad t]$

$$\begin{aligned}\boldsymbol{\varepsilon}_{j,t_j} &= \frac{1}{2}\int_0^{t_j}\left((\mathbf{g}_{j-1}-\mathbf{g}^*)\times + \frac{1}{2}(\mathbf{g}_{j-1}\mathbf{g}_{j-1}^T - \mathbf{g}^*\mathbf{g}^{*T})\right)\boldsymbol{\omega}\,dt \\ &= \frac{1}{2}\int_0^{t_j}\left(\boldsymbol{\varepsilon}_{j-1}\times + \frac{1}{2}(\boldsymbol{\varepsilon}_{j-1}\mathbf{g}_{j-1}^T + \mathbf{g}^*\boldsymbol{\varepsilon}_{j-1}^T)\right)\boldsymbol{\omega}\,dt \\ &= \frac{1}{2}t_j\left[\boldsymbol{\varepsilon}_{j-1,t_{j-1}}\times + \frac{1}{2}(\boldsymbol{\varepsilon}_{j-1,t_{j-1}}\mathbf{g}_{j-1,t_{j-1}}^T + \mathbf{g}^*_{t_{j-1}}\boldsymbol{\varepsilon}_{j-1,t_{j-1}}^T)\right]\boldsymbol{\omega}_{t_{j-1}} \\ &\triangleq \frac{1}{2}\left[\boldsymbol{\varepsilon}_{j-1,t_{j-1}}\times + \frac{1}{2}(\boldsymbol{\varepsilon}_{j-1,t_{j-1}}\mathbf{g}_{j-1,t_{j-1}}^T + \mathbf{g}^*_{t_{j-1}}\boldsymbol{\varepsilon}_{j-1,t_{j-1}}^T)\right]\boldsymbol{\theta}_{t_{j-1}} \\ &= \frac{1}{2}\left[-\boldsymbol{\theta}_{t_{j-1}}\times + \frac{1}{2}(\mathbf{g}_{j-1,t_{j-1}}^T\boldsymbol{\theta}_{t_{j-1}}\mathbf{I}_3 + \mathbf{g}^*_{t_{j-1}}\boldsymbol{\theta}_{t_{j-1}}^T)\right]\boldsymbol{\varepsilon}_{j-1,t_{j-1}}\end{aligned}$$
(21)

where $t_{j-1} \in (0 \quad t_j)$ and $\boldsymbol{\theta}_{t_{j-1}} \triangleq t_j \boldsymbol{\omega}_{t_{j-1}}$. The third equality is valid by the Mean-value Theorem of integrals. Repeat the above process, and for the considerably small time interval $[0 \quad t]$ where $\mathbf{g}$ is a small quantity

$$\begin{aligned}\boldsymbol{\varepsilon}_{j,t_j} &= \frac{1}{2^j}\prod_{k=1}^j\left[-\boldsymbol{\theta}_{t_{k-1}}\times + \frac{1}{2}(\mathbf{g}_{k-1,t_{k-1}}^T\boldsymbol{\theta}_{t_{k-1}}\mathbf{I}_3 + \mathbf{g}^*_{t_{k-1}}\boldsymbol{\theta}_{t_{k-1}}^T)\right]\boldsymbol{\varepsilon}_{0,t_0} \\ &= \ldots = O\left(\frac{1}{2^j}\left(((\boldsymbol{\varepsilon}_{0,t_0}\times\boldsymbol{\theta}_{t_0})\times\boldsymbol{\theta}_{t_1}\cdots)\times\boldsymbol{\theta}_{t_{j-2}}\right)\times\boldsymbol{\theta}_{t_{j-1}}\right)\end{aligned}$$
(22)

which means $|\boldsymbol{\varepsilon}_{j,t_j}| < O\left(\dfrac{|\boldsymbol{\varepsilon}_{0,t_0}|\prod_{k=0}^{j-1}|\boldsymbol{\theta}_{t_k}|}{2^j}\right)$. We see that if $\prod_{k=0}^\infty|\boldsymbol{\theta}_{t_k}|$ is bounded, then $\boldsymbol{\varepsilon}_{j,t_j} \to 0$ as $j\to\infty$. So the iterative process (19) converges to the true Rodrigues vector function. ∎

*Proposition 1*: Given the true angular velocity function $\boldsymbol{\omega}$, the iterative process (19) converges to the true Rodrigues vector function over the interval when $t\sup|\boldsymbol{\omega}| < 2$.

Proof. The boundedness condition in *Theorem 1* can be further relaxed,
$$\prod_{k=0}^{j-1}|\boldsymbol{\theta}_{t_k}| = \prod_{k=0}^{j-1}|\boldsymbol{\omega}_{t_k}|t_{k+1} < (\sup|\boldsymbol{\omega}|)^j\prod_{k=0}^{j-1}t_{k+1} < (t\sup|\boldsymbol{\omega}|)^j$$
(23)

where $\sup(\cdot)$ denotes the function supreme value over the time interval of interest. Equation (23) means $|\boldsymbol{\varepsilon}_{j,t_j}| < O(|\boldsymbol{\varepsilon}_{0,t_0}|(t\sup|\boldsymbol{\omega}|/2)^j)$. Therefore, if $t\sup|\boldsymbol{\omega}| < 2$, $\boldsymbol{\varepsilon}_{j,t_j} \to 0$ as $j\to\infty$. ∎

*Proposition 1* provides a looser criterion that is easier to check the boundedness condition, namely, the product of the time interval length and the supreme magnitude of the angular velocity function over the time interval is less than 2. It can also be used to roughly determine the iteration times required to achieve the desirable accuracy, e.g., for a desirable accuracy $\delta$, we need $|\boldsymbol{\varepsilon}_{0,t_0}|(t\sup|\boldsymbol{\omega}|/2)^j < \delta$ or equivalently the iteration times

$$j > \frac{\ln(\delta/|\boldsymbol{\varepsilon}_{0,t_0}|)}{\ln(t\sup|\boldsymbol{\omega}|/2)}$$
(24)

Note the above analysis can alternatively be recast in the context of general differential equations, for which the Picard–Lindelöf theorem guarantees the iteration process's existence and uniqueness [20]. However, by making use of the specific form of the Rodrigues vector differential equation (14), *Theorem 1* gives more delicate result than the general conclusion of the Picard–Lindelöf theorem.

In practice, the angular velocity function is not perfect and always contains some kinds of errors. The following theorem describes what happens to the iteration process with an error-contaminated angular velocity function $\hat{\boldsymbol{\omega}} = \boldsymbol{\omega} + \boldsymbol{\delta}$, where $\boldsymbol{\delta}$ denotes the error function.

*Theorem 2*: Given an error-contaminated angular velocity function $\hat{\boldsymbol{\omega}} = \boldsymbol{\omega} + \boldsymbol{\delta}$, the iterative process (19) converges to the Rodrigues vector function that corresponds to the error-contaminated angular velocity function, when $\prod_{k=0}^\infty|\hat{\boldsymbol{\theta}}_{t_k}|$ is bounded.

The proof is straightforward by referring to that of *Theorem 1*.

*Proposition 2*: Given an error-contaminated angular velocity function $\hat{\boldsymbol{\omega}} = \boldsymbol{\omega} + \boldsymbol{\delta}$, the discrepancy between the convergence result of the iterative process (19) and the true Rodrigues vector function is less than $t\sup|\boldsymbol{\delta}|$ in magnitude if the time interval is considerably small.

Proof. Similar with *Theorem 1*, for the iterative process (19) using the error-contaminated angular velocity function $\hat{\boldsymbol{\omega}} = \boldsymbol{\omega} + \boldsymbol{\delta}$, the error function $\boldsymbol{\varepsilon}_j$ at time $t_j \in [0 \quad t]$

$$\begin{aligned}\boldsymbol{\varepsilon}_{j,t_j} &= \boldsymbol{\delta}_{\theta,t_{j-1}} + \frac{1}{2}\left[-\hat{\boldsymbol{\theta}}_{t_{j-1}}\times + \frac{1}{2}(\mathbf{g}_{j-1,t_{j-1}}^T\hat{\boldsymbol{\theta}}_{t_{j-1}}\mathbf{I}_3 + \mathbf{g}^*_{t_{j-1}}\hat{\boldsymbol{\theta}}_{t_{j-1}}^T)\right]\boldsymbol{\varepsilon}_{j-1,t_{j-1}} \\ &\triangleq \mathbf{A}_{j-1} + \frac{1}{2}\mathbf{B}_{j-1}\boldsymbol{\varepsilon}_{j-1,t_{j-1}} \\ &= \mathbf{A}_{j-1} + \frac{1}{2}\mathbf{B}_{j-1}\left[\mathbf{A}_{j-2} + \frac{1}{2}\mathbf{B}_{j-2}\right]\boldsymbol{\varepsilon}_{j-2,t_{j-2}} \\ &= \mathbf{A}_{j-1} + \frac{1}{2}\mathbf{B}_{j-1}\mathbf{A}_{j-2} + \frac{1}{2^2}\prod_{k=j-1}^j\mathbf{B}_{k-1}\boldsymbol{\varepsilon}_{j-2,t_{j-2}} \\ &=\ldots= \mathbf{A}_{j-1} + \frac{1}{2}\prod_{k=j}^j\mathbf{B}_{k-1}\mathbf{A}_{j-2} + \frac{1}{2^2}\prod_{k=j-1}^j\mathbf{B}_{k-1}\mathbf{A}_{j-3} \\ &\quad +\ldots+ \frac{1}{2^{j-1}}\prod_{k=2}^j\mathbf{B}_{k-1}\mathbf{A}_0 + \frac{1}{2^j}\prod_{k=1}^j\mathbf{B}_{k-1}\boldsymbol{\varepsilon}_{0,t_0} \\ &= \boldsymbol{\delta}_{\theta,t_{j-1}} + O\left(\frac{1}{2}\boldsymbol{\delta}_{\theta,t_{j-2}}\times\hat{\boldsymbol{\theta}}_{t_{j-1}}\right) + O\left(\frac{1}{2^2}(\boldsymbol{\delta}_{\theta,t_{j-3}}\times\hat{\boldsymbol{\theta}}_{t_{j-2}})\times\hat{\boldsymbol{\theta}}_{t_{j-1}}\right) \\ &\quad +\ldots+ O\left(\frac{1}{2^j}(((\boldsymbol{\varepsilon}_{0,t_0}\times\hat{\boldsymbol{\theta}}_{t_0})\times\hat{\boldsymbol{\theta}}_{t_1}\cdots)\times\hat{\boldsymbol{\theta}}_{t_{j-2}})\times\hat{\boldsymbol{\theta}}_{t_{j-1}}\right)\end{aligned}$$
(25)

where $\boldsymbol{\delta}_{\theta,t_{j-1}} \triangleq t_j\boldsymbol{\delta}_{t_{j-1}}$ and $\hat{\boldsymbol{\theta}}_{t_{j-1}} \triangleq t_j\hat{\boldsymbol{\omega}}_{t_{j-1}}$. For a considerably small time interval $[0 \quad t]$, as $\hat{\boldsymbol{\theta}}$ and $\mathbf{g}$ are small quantities,

$$\varepsilon_{j,t_j} = O(\delta_{\theta,t_{j-1}}) + O\left(\frac{1}{2^j}\left(\left(\left(\varepsilon_{0,t_0}\times\hat{\theta}_{t_0}\right)\times\hat{\theta}_{t_1}\cdots\right)\times\hat{\theta}_{t_{j-2}}\right)\times\hat{\theta}_{t_{j-1}}\right) \quad (26)$$

We see that as $j \to \infty$, the second term approaches to zero and $|\varepsilon_{j,t_j}| = O(|t_j \delta_{t_{j-1}}|) < t\sup|\delta|$. ∎

Proposition 2 actually states an intuitive fact that the iterative computation accuracy of the Rodrigues vector is limited by the quality of the provided angular velocity function, which may be adversely affected by the gyroscope measurement errors and the fitting method. From another perspective, (26) depicts the quantitative contribution of both measurement errors and computation errors to the attitude reconstruction imperfection. In practice, the latter should reasonably be kept less than the former. With *Proposition 1*, it means

$$t\sup|\delta| > |\varepsilon_{0,t_0}|\left(t\sup|\hat{\omega}|/2\right)^j, \text{ i.e., } j > \frac{\ln\left(t\sup|\delta|/|\varepsilon_{0,t_0}|\right)}{\ln\left(t\sup|\hat{\omega}|/2\right)} \quad (27)$$

*B. Angular Velocity Polynomial Function Fitted from Gyroscope Measurements*

In practice, we cannot directly get the angular velocity function, but are only accessible to the error-contaminated discrete measurements of angular velocity or angular increment, e.g., by gyroscopes. In view of the fact that the Chebyshev polynomial has better numerical stability than the normal polynomial [21], we use the Chebyshev polynomial to fit the discrete angular velocity or angular increment measurements. The Chebyshev polynomial of the first kind is defined by the recurrence relation as [21]

$$F_0(x) = 1, F_1(x) = x, F_{j+1}(x) = 2xF_j(x) - F_{j-1}(x) \quad (28)$$

where $F_i(x)$ is the $i^{th}$-degree Chebyshev polynomial of the first kind.

Suppose $N$ angular velocity/increment measurements are used. The considered time interval is $[0 \; t_N]$. In order to use the Chebyshev polynomials, it is required to map the time interval to $[-1 \; 1]$. Let $t = \frac{t_N}{2}(1+\tau)$, (19) becomes

$$\mathbf{g}_{j+1} = \int_0^t \left(\mathbf{I}_3 + \frac{1}{2}\mathbf{g}_j\times + \frac{1}{4}\mathbf{g}_j\mathbf{g}_j^T\right)\boldsymbol{\omega}\,dt$$
$$= \frac{t_N}{2}\int_{-1}^{\tau}\left(\mathbf{I}_3 + \frac{1}{2}\mathbf{g}_j\times + \frac{1}{4}\mathbf{g}_j\mathbf{g}_j^T\right)\boldsymbol{\omega}\,d\tau \quad (29)$$

For $N$ angular velocity measurements $\tilde{\boldsymbol{\omega}}_{t_k}, k = 1, 2, \ldots N$ over the time interval $[0 \; t_N]$, the angular velocity can be fitted by the Chebyshev polynomial in time up to the order of $N$-1,

$$\hat{\boldsymbol{\omega}} = \sum_{i=0}^{n} \mathbf{c}_i F_i(\tau), \; n \leq N-1 \quad (30)$$

where the coefficients $\mathbf{c}_i$ can be determined by solving the equation

$$\mathbf{A}_\omega \begin{bmatrix}\mathbf{c}_0^T \\ \mathbf{c}_1^T \\ \vdots \\ \mathbf{c}_n^T\end{bmatrix} \triangleq \begin{bmatrix} 1 & F_1(\tau_1) & \cdots & F_n(\tau_1) \\ 1 & F_1(\tau_2) & \cdots & F_n(\tau_2) \\ \vdots & \vdots & \vdots & \vdots \\ 1 & F_1(\tau_N) & \cdots & F_n(\tau_N) \end{bmatrix} \begin{bmatrix}\mathbf{c}_0^T \\ \mathbf{c}_1^T \\ \vdots \\ \mathbf{c}_n^T\end{bmatrix} = \begin{bmatrix}\tilde{\boldsymbol{\omega}}_{t_1}^T \\ \tilde{\boldsymbol{\omega}}_{t_2}^T \\ \vdots \\ \tilde{\boldsymbol{\omega}}_{t_N}^T\end{bmatrix} \quad (31)$$

Similarly, for $N$ angular increment measurements $\Delta\tilde{\boldsymbol{\theta}}_{t_k}, k = 1, 2, \ldots N$, the angular velocity can also be fitted by the Chebyshev polynomial in time up to the order of $N$-1. According to the integral property of the Chebyshev polynomial [21],

$$G_{i,[\tau_{k-1}\tau_k]} \triangleq \int_{\tau_{k-1}}^{\tau_k} F_i(\tau)d\tau = \begin{cases}\left(\frac{iF_{i+1}(\tau_k)}{i^2-1} - \frac{\tau_k F_i(\tau_k)}{i-1}\right) - \left(\frac{iF_{i+1}(\tau_{k-1})}{i^2-1} - \frac{\tau_{k-1}F_i(\tau_{k-1})}{i-1}\right), & i \neq 1 \\ \frac{\tau_k^2 - \tau_{k-1}^2}{2}, & i = 1\end{cases} \quad (32)$$

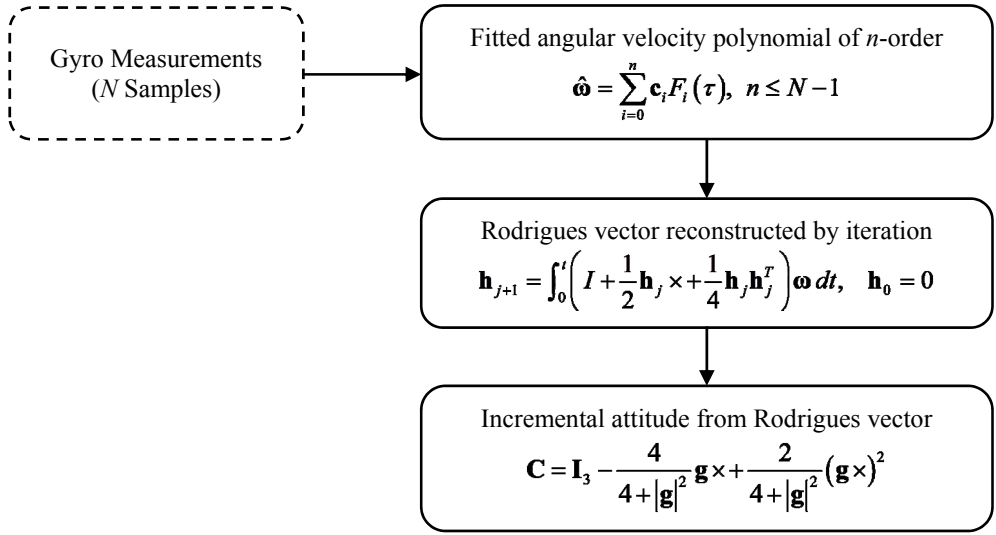

Figure 1. Flowchart of Rodrigues vector reconstruction and attitude update.

The fitted angular increment is related to the fitted angular velocity by

$$\Delta \hat{\boldsymbol{\theta}}_{t_k} = \int_{t_{k-1}}^{t_k} \hat{\boldsymbol{\omega}} dt = \frac{t_N}{2} \int_{\tau_{k-1}}^{\tau_k} \hat{\boldsymbol{\omega}} d\tau = \frac{t_N}{2} \int_{\tau_{k-1}}^{\tau_k} \sum_{i=0}^{n} \mathbf{c}_i F_i(\tau) d\tau$$
$$= \frac{t_N}{2} \sum_{i=0}^{n} \mathbf{c}_i \int_{\tau_{k-1}}^{\tau_k} F_i(\tau) d\tau = \frac{t_N}{2} \sum_{i=0}^{n} \mathbf{c}_i G_{i,[\tau_{k-1} \tau_k]}$$
(33)

Then the coefficients $\mathbf{c}_i$ can be determined by solving the equation

$$\mathbf{A}_\theta \begin{bmatrix} \mathbf{c}_0^T \\ \mathbf{c}_1^T \\ \vdots \\ \mathbf{c}_n^T \end{bmatrix} \triangleq \begin{bmatrix} G_{0,[\tau_0 \tau_1]} & G_{1,[\tau_0 \tau_1]} & \cdots & G_{n,[\tau_0 \tau_1]} \\ G_{0,[\tau_1 \tau_2]} & G_{1,[\tau_1 \tau_2]} & \cdots & G_{n,[\tau_1 \tau_2]} \\ \vdots & \vdots & \vdots & \vdots \\ G_{0,[\tau_{N-1} \tau_N]} & G_{1,[\tau_{N-1} \tau_N]} & \cdots & G_{n,[\tau_{N-1} \tau_N]} \end{bmatrix} \begin{bmatrix} \mathbf{c}_0^T \\ \mathbf{c}_1^T \\ \vdots \\ \mathbf{c}_n^T \end{bmatrix} = \frac{2}{t_N} \begin{bmatrix} \Delta \tilde{\boldsymbol{\theta}}_{t_1}^T \\ \Delta \tilde{\boldsymbol{\theta}}_{t_2}^T \\ \vdots \\ \Delta \tilde{\boldsymbol{\theta}}_{t_N}^T \end{bmatrix}$$
(34)

*C. Rodrigues Vector Reconstruction from Fitted Angular Velocity Polynomial*

Figure 1 plots the flowchart of how to reconstruct the Rodrigues vector from gyroscope measurements and to update the attitude. With the boundedness condition in mind, substituting the fitted angular velocity function (30) into the iteration process (29) is supposed to well reconstruct the Rodrigues vector function as a polynomial, as the group of polynomials are closed under elementary arithmetic operations. This is an appreciated benefit brought about by the simplicity of the Rodrigues vector's differential function (14). Referring to (29), we can see that the polynomial order of the Rodrigues vector at current iteration is equal to two times the order of that at previous iteration plus the order of the angular velocity and 1, i.e., $order(\mathbf{g}_j) = 2 \times order(\mathbf{g}_{j-1}) + n + 1$ with $order(\mathbf{g}_0) = 0$.

Analytic development of the iterative process (29) is straightforward but tedious, and one may turn to a symbolic computation toolbox in Matlab to spare the laborious work. The obtained polynomial expression can be readily saved as a function to hand.

As the Chebyshev polynomial has the same theoretical result with the normal polynomial, for better comparison we use the latter instead to exemplify the explanation. Take $N = 2, n = 1$ as an example, the fitted angular velocity function $\hat{\boldsymbol{\omega}} = \mathbf{c}_0 + \mathbf{c}_1 t$. Table I lists the reconstructed Rodrigues vector polynomials for the first two iterations. The polynomial order is 2 at the first iteration, 6 at the second iteration, and will be 14 for another iteration. Larger number of samples ($N$) or more iteration ($j$) yields a higher order Rodrigues vector polynomial.

*D. Approximate Rotation Vector Reconstruction from Fitted Angular Velocity Polynomial*

Approximate the rotation vector differential equation (10) as

$$\dot{\mathbf{g}} \approx \boldsymbol{\omega} + \frac{1}{2} \mathbf{g} \times \boldsymbol{\omega} + \frac{1}{12} \mathbf{g} \times (\mathbf{g} \times \boldsymbol{\omega}) \quad (35)$$

In principle, an iteration process similar to (19) could also be used to reconstruct the approximate rotation vector (named RotFIter), i.e.,

$$\mathbf{g}_{j+1} = \int_0^t \left( \mathbf{I}_3 + \frac{1}{2} \mathbf{g}_j \times + \frac{1}{12} (\mathbf{g}_j \times)^2 \right) \boldsymbol{\omega} \, dt \quad (36)$$

with $\mathbf{g}_0 = 0$. It should be highlighted that this iteration process would not converge to the rotation vector, but some kind of vector whose rate equation is exactly represented by (35). It can be readily checked that two iterations will produce (17) in the integral form. In the mainstream attitude algorithm with $N = 2$, the approximate rotation vector that further discards the third term in (36) is [13]

$$\mathbf{g} \approx \int_0^t \left( \mathbf{I}_3 + \frac{1}{2} \left( \mathbf{c}_0 t + \frac{1}{2} \mathbf{c}_1 t^2 \right) \times \right) (\mathbf{c}_0 + \mathbf{c}_1 t) \, dt = t \mathbf{c}_0 + \frac{t^2}{2} \mathbf{c}_1 + \frac{t^3}{12} \mathbf{c}_0 \times \mathbf{c}_1$$
(37)

which lacks all the triple products of coefficients $\mathbf{c}_i$, in contrast to the Rodrigues vector polynomial at the second iteration in Table II. The mainstream attitude algorithms for more (angular-increment) samples have been discussed in [2, 5, 15].

IV. EXTENSION TO GENERAL RIGID MOTION

The Chasles theorem, in analogy to the Euler theorem for pure rotation, states that the general motion of a rigid body in space consists of a rotation about an axis and a translation parallel to that axis [22]. Among many mathematical representations, dual quaternion is the most compact and efficient way to express the general rigid motion [23-25]. In fact, dual quaternion shares many common properties with quaternion. According to the principle of transference by Kotelnikov [26], the characteristics of quaternion are completely inherited by dual quaternion. Therefore the above analyses and conclusions obtained from rotation/quaternion naturally extend to the general motion/dual quaternion. For example, readers are referred to [27] for an application of dual quaternion in inertial navigation representation and computation. Therein the inertial navigation principle is formulated by three dual quaternion differential equations of identical form with the quaternion differential equation (2), so the attitude reconstruction method proposed in this paper can be readily applied to the general motion problem.

Table II. Rodrigues Vector Polynomials for Two Iterations ($N = 2, n = 1$)

| Iteration | Rodrigues Vector Polynomial | Order of Polynomial |
|---|---|---|
| j=1 | $\mathbf{g}_1 = \mathbf{c}_0 t + \frac{1}{2} \mathbf{c}_1 t^2$ | 2 |
| j=2 | $\mathbf{g}_2 = t\mathbf{c}_0 + \frac{t^2}{2} \mathbf{c}_1 + \frac{t^3}{12} \left( \mathbf{c}_0 \times \mathbf{c}_1 + \mathbf{c}_0 \mathbf{c}_0^T \mathbf{c}_0 \right) + \frac{t^4}{32} \left( \mathbf{c}_1 \mathbf{c}_0^T \mathbf{c}_0 + 3 \mathbf{c}_0 \mathbf{c}_0^T \mathbf{c}_1 \right)$ $+ \frac{t^5}{80} \left( 2 \mathbf{c}_0 \mathbf{c}_1^T \mathbf{c}_1 + 3 \mathbf{c}_1 \mathbf{c}_0^T \mathbf{c}_1 \right) + \frac{t^6}{96} \mathbf{c}_1 \mathbf{c}_1^T \mathbf{c}_1$ | 6 |

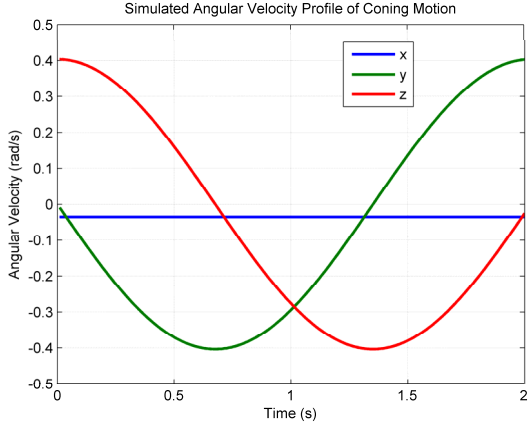

Figure 2. Simulated angular velocity of coning motion.

## V. SIMULATION RESULTS

In this section, the Rodrigues vector (and the corresponding attitude) is reconstructed and evaluated under the coning motion. The coning motion plays an important role in the inertial navigation field, as it has an analytic expression in the angular velocity and the associated generalized kinematic vector. Almost all attitude algorithms have been designed using the coning motion or its variants as the design criterion [6, 7, 10]. Note that the coning motion is used only as a testing motion scenario and has no affect on the design process of the RodFIter method.

The angular increment measurement from gyroscopes is assumed. The discrete angular increment measurements are uniformly generated at 100 Hz, i.e., the sampling time interval $T = 0.01s$. The following angle metric is used to quantify the attitude computation error

$$\varepsilon_{att} = 2\left\|\left[\mathbf{q}^* \circ \hat{\mathbf{q}}\right]_{2:4}\right\| \qquad (38)$$

where $\hat{\mathbf{q}}$ denotes the quaternion estimate from the reconstructed generalized kinematic vector, and $[\cdot]_{2:4}$ is the sub-vector formed by the last three elements of error quaternion. The angular velocity of the coning motion is given by

$$\boldsymbol{\omega} = \Omega\begin{bmatrix} -2\sin^2(\alpha/2) & -\sin(\alpha)\sin(\Omega t) & \sin(\alpha)\cos(\Omega t) \end{bmatrix}^T \qquad (39)$$

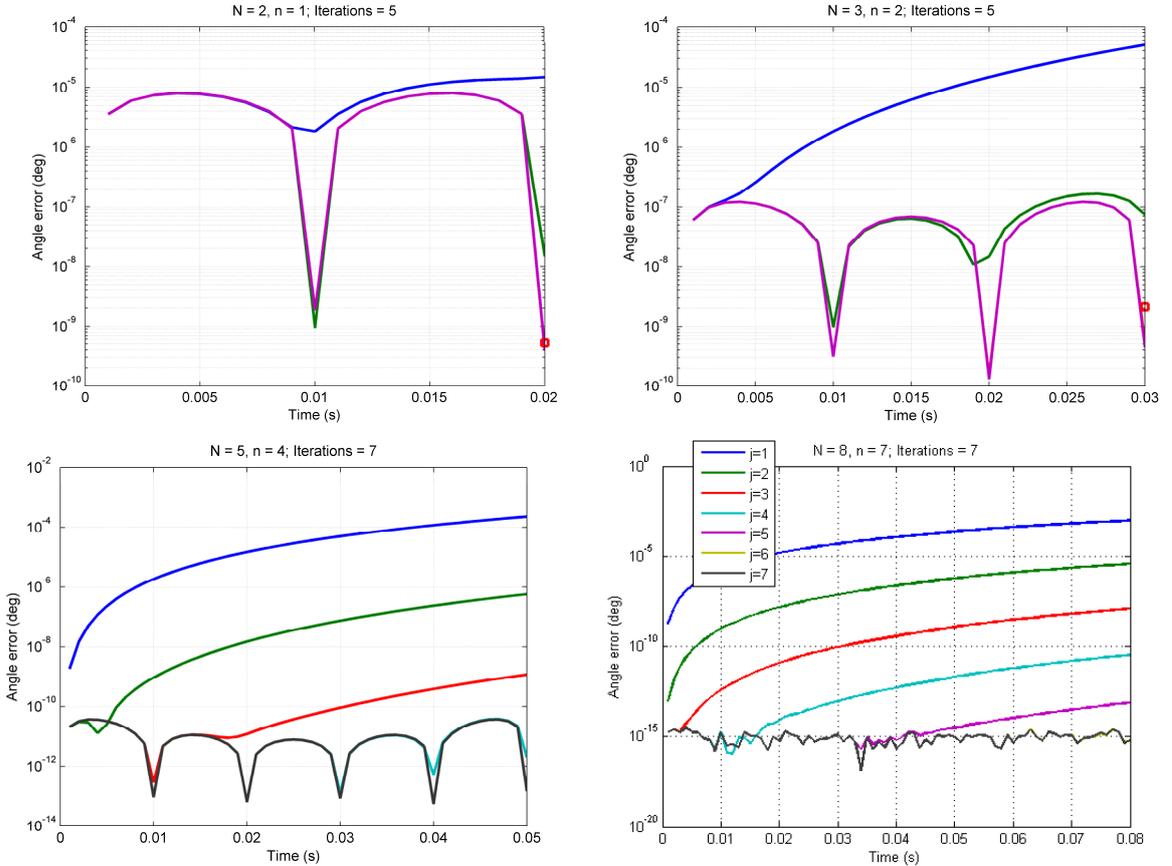

Figure 3. Attitude computation errors for coning motion ($N$ = 2, 3, 5 and 8). Red squares denote attitude error of mainstream algorithm ($N$ = 2 and 3) in (37).

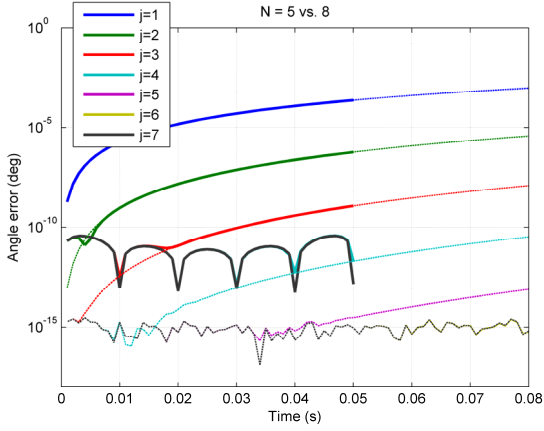

Figure 4. Combined plot of attitude errors for coning motion ($N = 5$, solid line; $N=8$, dashed line).

where the coning angle is set to $\alpha = 10 \deg$, the coning frequency $\Omega = 0.74\pi$. Figure 2 plots the angular velocity profile for the first two seconds. The magnitude of rotation is $\alpha$ and the unit vector of the fixed rotation axis $\mathbf{e} = \begin{bmatrix} 0 & \cos(\Omega t) & \sin(\Omega t) \end{bmatrix}^T$. It can be readily checked that the magnitude of $\boldsymbol{\omega}$ is time-invariant and $\mathbf{e}^T \boldsymbol{\omega} = 0$ which maximizes the noncommutativity rate terms of any generalized kinematic vector (6). This is the reason that the coning motion is widely used as the attitude algorithm design criterion. The corresponding Rodrigues vector $\mathbf{g} = 2\tan\left(\dfrac{\alpha}{2}\right)\mathbf{e}$, and the attitude quaternion can be obtained by (13). The angular increment measurements are used as the input for fitting the angular velocity polynomial (30).

Figure 3 presents the attitude computation errors for $N = 2, 3, 5$ and $8$, for which the order of the fitted angular velocity polynomials is all set to $n = N - 1$. The number of iteration is set to 5 for the former two cases, and 7 for the latter two. The errors generally decrease as the iteration number increases. It is observed that for the $N$-sample computation, more iterations after $N$ times cannot further improve the accuracy. This is owed to the $N$-order polynomials' insufficient approximation of the trigonometric functions in the angular velocity (39), while the flat error curve in the bottom-right subfigure for $N = 8$ indicates that the eight-order polynomials are quite sufficient in this case. This observation is confirmed when the angle errors of different $N$-s are put together, e.g., see Fig. 4 for the combined plot of $N = 5$ and 8. In principle, higher accuracy can be achieved by using a higher-order polynomial fitting of the angular velocity. In other words, more accurate the angular velocity approximation, more accurate is the attitude computation. Combined this observation with (27), the number of samples for fitting the angular velocity can be empirically selected to satisfy $N \approx j > \ln\left(t \sup |\boldsymbol{\delta}| / |\boldsymbol{\varepsilon}_{0,t_0}|\right) / \ln\left(t \sup |\boldsymbol{\omega}| / 2\right)$.

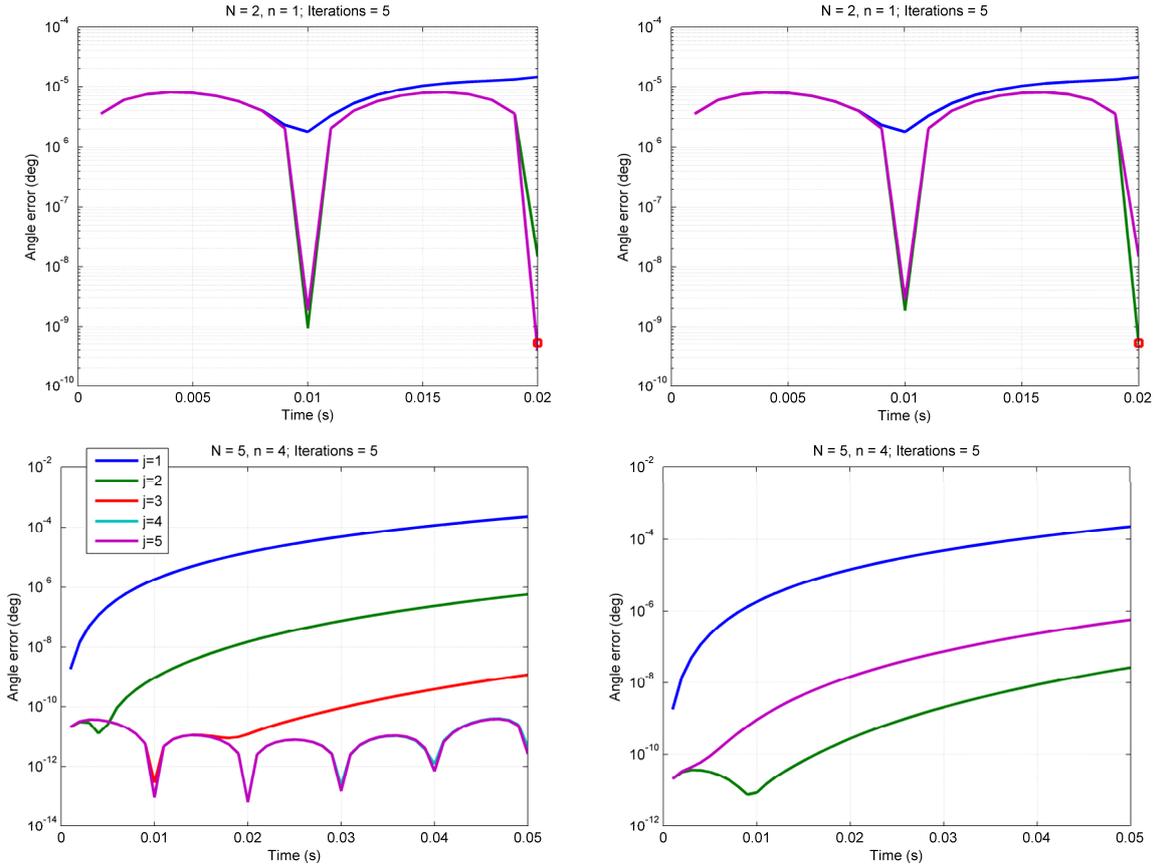

Figure 5. Attitude computation result for coning motion using approximate rotation vector by (36) (left column: RotFIter-T3 with the third term; right column: RotFIter-T2 without the third term). Red square denotes attitude error of mainstream algorithm ($N = 2$) in (37).

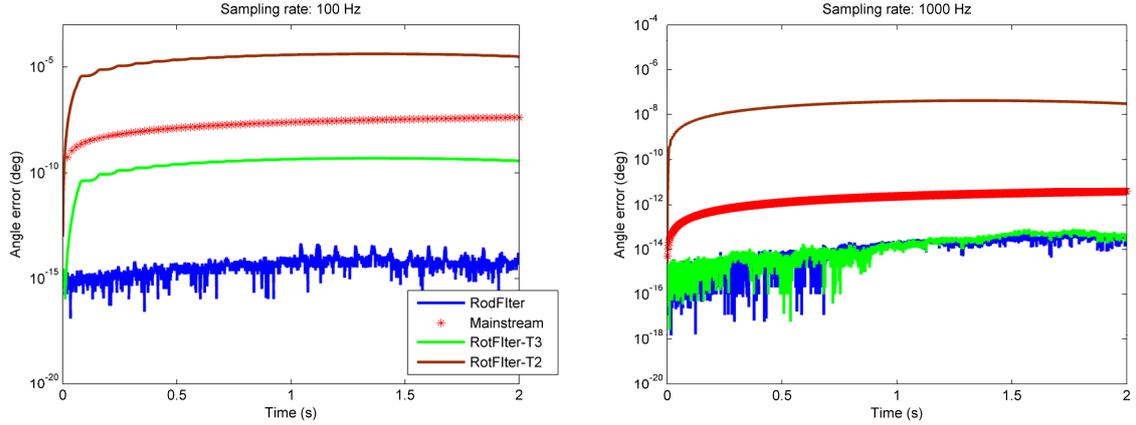

Figure 6. Attitude errors comparison of RodFIter method ($N = 8$), rotation vector reconstruction with the third term (RotFIter-T3, $N = 8$) and without the third term (RotFIter-T2, $N = 8$), and mainstream algorithm ($N = 2$). (Left sub-figure: 100 Hz; right sub-figure: 1000 Hz)

The attitude error of the mainstream algorithms for $N = 2$ (c.f. (37)) and $N = 3$ (c.f. [5]) are also given as a comparison (red squares in the upper two subfigures, Fig. 3). Regardless of sensor imperfection, the RodFIter method improves the attitude accuracy by about seven orders of magnitude over the mainstream algorithms (the lower-right subfigure for $N = 8$ versus the upper two subfigures for $N = 2, 3$). This extraordinary accuracy benefit is guaranteed by the proven convergence in *Theorem 1*. Note that all of the mainstream algorithms using $N$ samples can only yield an attitude result at the time of the $N$-th sample, while the RodFIter method is analytic and able to produce attitude results over the whole iteration time interval.

Figure 5 plots the attitude computation result using the approximate rotation vector of the iteration process (36) for $N = 2$ and 5, with the third term (denoted as RotFIter-T3) or without the third term (denoted as RotFIter-T2). The upper-right subfigure reveals how the mainstream algorithm (37) for $N = 2$ behaves over the whole time interval for the first time, while the upper-left subfigure shows the positive effect of incorporating the third term. The attitude accuracy at the end of the interval is improved by over one order. In contrast to the Rodrigues vector result (lower-left subfigure in Fig. 3), however, the accuracy of RotFIter-T2 cannot be further improved after two iterations. When the third term is discarded in RotFIter-T2, the result at the second iteration is unexpectedly the best and begins to degrade at further iterations (lines for the 3th-5th iterations overlapped). This phenomenon occurs because the iteration (36) converges to the some other vector whose rate equation is exactly represented by the first two terms, instead of the true rotation vector.

The left sub-figure of Figure 6 compares the attitude computation errors over the first two seconds of the RodFIter method ($N = 8$) and the approximate rotation vector reconstruction (RotFIter, $N = 8$) and the mainstream algorithm ($N = 2$). The number of iteration is set to 6. The RodFIter method yields the most accurate and almost drift-less attitude result. The approximate rotation vector reconstruction with the third term (RotFIter-T3) comes at the second best. Astonishingly, the approximate rotation vector reconstruction without the third term (RotFIter-T2) performs the worst, even worse than the mainstream algorithm ($N = 2$). It explains a frequently-encountered but never-answered phenomenon that using more samples based on the unduly-simplified rotation vector rate equation leads to worse rather than better accuracy. The right sub-figure of Figure 6 repeats the above comparison except that the sampling rate is increased to 1000 Hz. The errors of RotFIter-T3, RotFIter-T2 and mainstream algorithms ($N = 2$) are remarkably reduced by several orders. Notably, the RotFIter-T3 accuracy becomes comparable to the RodFIter. In contrast, the RodFIter method, still the best, gains little from the increased sampling rate, which implies the motion-incurred computation error has been completely suppressed at the 100 Hz sampling rate. This observation highlights the striking capability of the RodFIter method in mitigating computation errors.

As the angular velocity/increment measurements are unavoidably contaminated by errors, sensor bias ($\begin{bmatrix} 5 & -3 & 4 \end{bmatrix}^T \times 10^{-3}$ deg/h) and stochastic noise (0.002 $\deg/\sqrt{h}$) are added to the gyroscope measurements to examine the error sensitivity of the RodFIter method. Figure 7 plots the attitude error of an implementation with error-contaminated measurements. The RodFIter method stabilizes within two iterations, and the result of $N = 5$ is much better especially between the sample times, e.g., at t = 0.015s. The error of the mainstream algorithm (37) at t = 0.02s is comparable to that of the RodFIter method, which indicates that the sensor error dominates the attitude computation error for the above simulation setting. The coning frequency is then set to $\Omega = 10\pi$ so as to intentionally enlarge the adverse effect of high manoeuvrs on the computation error. As shown in Fig. 8, more samples are significantly helpful to the RodFIter method in reducing the attitude errors because the computation error

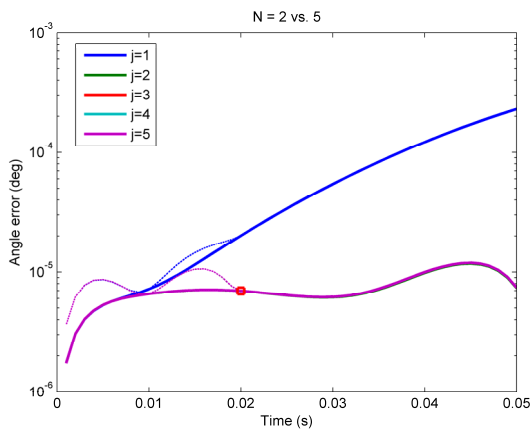

Figure 7. Attitude computation result for coning motion considering sensor errors ($N = 2$, dashed line; $N = 5$, solid line). Red square denotes attitude error of mainstream algorithm ($N = 2$) in (37).

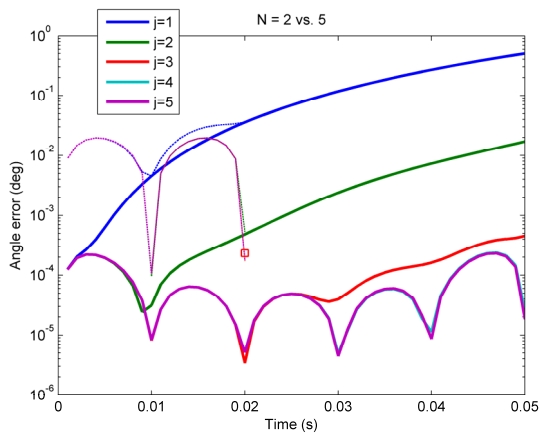

Figure 8. Attitude computation result for severe coning motion $\Omega = 10\pi$ considering sensor errors ($N = 2$, dashed line; $N = 5$, solid line). Red square denotes attitude error of mainstream algorithm ($N = 2$) in (37).

prevails over the sensor error in this case.

According to *Proposition 1*, a sufficient convergence precondition of the RodFIter method is $t \sup|\omega| < 2$. We have done additional tests to explore the practical convergence region from two aspects: the supreme angular velocity and the number of samples (equivalently the iterative integration time length $t = N \times T$). Throughout the tests, the iteration times are constantly set to seven. Specifically, we enlarge two simulation parameters, namely, the coning frequency $\Omega$ and the number of samples $N$, to see when the RodFIter method becomes to diverge, and mark the corresponding $\begin{bmatrix} N & \sup|\omega| \end{bmatrix}$ as the boundary of the practical convergence region. The maximum samples is ten. Figure 9 plots the test result together with the theoretical condition $t \sup|\omega| < 2$. The practical convergence region (area formed by the line and the x-y positive axes) is roughly consistent with predicted. It is obtained for fixed $T = 0.01s$. That the maximum angular velocity magnitude stays below 140 rad/s for $N = 2$ does not mean the RodFIter method cannot be used for motion with larger rotation rate. As the signal frequency in (39) is $\Omega/2\pi$, the sampling rate $1/T$ should be well above the Nyquist frequency $\Omega/\pi$ for a quality fitting of the angular velocity. In fact, the maximum admissible rotation rate could be raised by choosing a higher sampling rate. For example, for $T = 0.001s$, the RodFIter method does not diverge until the maximum angular velocity magnitude surpasses 1500 rad/s.

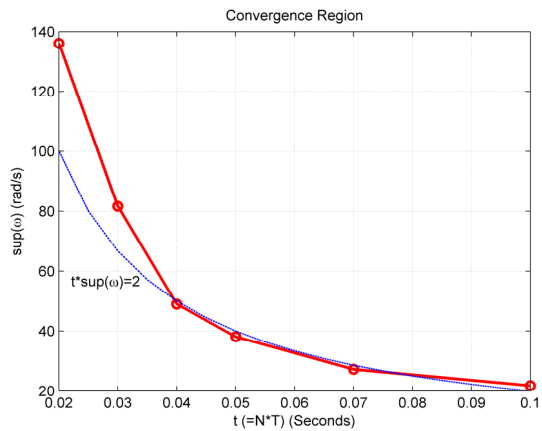

Figure 9. Convergence region boundary for fixed $T = 0.01s$: theoretical vs. practical. (Convergence region formed by the line and x-y positive axes)

As can be predicted, the RodFIter method imposes a huge computation load. Roughly quantified by the running time in Matlab, for example, the computation load of RodFIter ($N$=8) for 2-6 iterations is about 500-11000 times that of the mainstream algorithm ($N$=2). Hopefully, it would be alleviated by a more efficient design or hardware implementation.

## VI. CONCLUSIONS

The state-of-the-art attitude algorithms in the inertial navigation field have unexceptionally relied on the simplified differential equation of the rotation vector to compute the attitude. This paper raises the RodFIter method based on the analytic iterative integration, which is enabled by the Rodrigues vector's polynomial-like differential equation and the polynomial fitting of the angular velocity/increment measurements. It is proved that the reconstructed Rodrigues vector converges to the truth if the fitted angular velocity is exact. The mainstream algorithms using multiple samples can only yield an attitude result at the time of the final sample, while the RodFIter method is analytic and able to produce attitude results over the whole iteration time interval. Simulation tests under the attitude coning motion show that the proposed method produces a potentially ultimate attitude algorithm with the highest accuracy possible. The main idea of the RodFIter method naturally extends to the general rigid motion computation as a result of the resemblance between attitude/quaternion and general motion/dual quaternion. This work is believed having eliminated the long-standing theoretical barrier in exact motion integration from inertial measurements and diverted our attention completely to how to improve the inertial sensor quality and how to implement the proposed computation method efficiently.

ACKNOWLEDGEMENTS

Patent pending PCT/CN2017/082317. Thanks to Landis Markley for commenting on an early version and Qi Cai for mentioning the Mean-value Theorem.

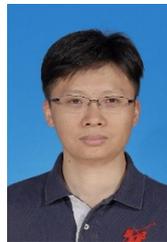

**Yuanxin Wu** received the B.Sc. and Ph.D. degree in navigation from Department of Automatic Control, National University of Defense Technology, in 1998 and 2005, respectively. He was with National University of Defense Technology as Lecturer (2006-2007) and Associate Professor (2008-2012), Department of Geomatics Engineering, University of Calgary, Canada, as a visiting Postdoctoral Fellow (2009.2-2010.2) and Central South

12University as Professor (2013-2015). He is currently a Professor of Navigation and Control in School of Electronic Information and Electrical Engineering, Shanghai Jiao Tong University, China. His current research interests include inertial-based navigation system, state estimation, inertial-visual fusion and wearable human motion sensing. He was the recipient of National Excellent Doctoral Dissertation (2008), New Century Excellent Talents in University (2010), Fok Ying Tung Education Fellowship (2012), NSFC Award for Excellent Young Scientists (2014), Natural Science and Technology Award in University (2008, 2016) and Elsevier's Most Cited Chinese Researchers (Aerospace Engineering, 2015-2017). He serves as an Associate Editor (2013-, 2016-) for The *Journal of Navigation* and *IEEE Trans. on Aerospace and Electronic Systems*.